\documentclass[reprint, aps,prd,twocolumn,showpacs,showkeys, 10pt, longbibliography, nolinenumbers, floatfix]{revtex4-1}

\usepackage{amsmath}
\usepackage{graphicx}
\usepackage{float}
\usepackage{dcolumn}
\usepackage{multirow}
\usepackage{natbib}
\usepackage{bm}
\usepackage{changes}
\usepackage{comment}
\usepackage[colorlinks = true,linkcolor = blue,urlcolor  = blue,citecolor = blue,anchorcolor = blue]{hyperref}
\usepackage{enumerate}
\usepackage{cleveref}

\begin{document}

\title{Binding energy of compact stars and their non-radial oscillations}

\author{P. Laskos-Patkos $^{1}$}
\email{plaskos@physics.auth.gr}

\author{S. Papadopoulos $^{2}$}

\author{Ch.C. Moustakidis $^{1}$}
\email{moustaki@auth.gr}

\affiliation{$^{1}$ Department of Theoretical Physics, Aristotle University of Thessaloniki, 54124 Thessaloniki, Greece}

\affiliation{$^{2}$ Département de Physique, Université Claude Bernard Lyon1, Lyon, 69622, France}

\begin{abstract}
In the past years, a significant effort has been made with the scope of determining correlations, involving compact star properties, that are independent of the nuclear equation of state. Such universal relations are of utmost importance as they allow for the imposition of constraints on stellar properties without directly measuring them and they may also serve as a probe of General Relativity. In the present study, we investigated the possible existence of a universal relation between the binding energy of compact stars and the frequency of their non-radial oscillations. The main motivation was related to the fact that both of the aforementioned quantities might be measured in the occurrence of a supernova explosion. Interestingly, we found that there is a empirical relation between the oscillation frequency and the binding energy for both $f$ and $p_1$ modes, assuming hadronic stellar matter. The inclusion of hybrid equations of state, incorporating sharp phase transitions, was shown to result into deviations from the aforementioned quasi-universal relation. 


\end{abstract}

\maketitle





\section{Introduction}
\label{introduction}
Neutron stars are considered to be unique laboratories for the study of dense nuclear matter and strong gravitational fields. In the past decades, a continuous research endeavor has allowed physicists to probe the nuclear equation of state and some of its properties via neutron star observations~\cite{Abbott-2018,Antoniadis-2013,Cromatie-2020,Romani-2022,Miller-2019,Riley-2019}, and also to examine the impact of considering modified theories of gravity on neutron star structure~\cite{Pani-2011,Orellana-2013,Abbas-2015,Das-2016}. {However, current limitations in the accuracy of astronomical measurements do not allow for an extraction of the exact form of the nuclear equation of state (EOS). In addition, certain EOSs that would be ruled out by neutron star observations could be rendered viable by the consideration of an alternative theory of gravity (e.g., larger maximum mass predictions, see ref.~\cite{Orellana-2013}). As a consequence, there is a degeneracy concerning the possible interpretation of astronomical data.}

In the past years, there has been extensive work on tackling issues resulting from the (currently) unknown nature of the nuclear EOS. A major step has been achieved by the proposition of universal or quasi-universal relations, which describe the dependence between the structural properties of a compact object in a way that is independent of the underlying nuclear model~\cite{Yagi-2013,Maselli-2013,Yagi-2015,Yagi-2017,Alexander-2019,Martinon-2014,Brew-2016,Chatterjee-2025,Jyothilakshmi-2025,Anderson-1998,Lau-2010,Chen-2014,Chirenti,Lioutas-2021,Zhao-2022,Rodriguez-2025,Lattimer-2001,Reed-2020,Laskos-2022,Marques-2017,Raduta-2020,Khadkikar-2021,Largani-2022,Ranea-2023,Ranea-2022,Ranea-2022b}.~{Such relations have been discussed for cold neutron stars~\cite{Yagi-2013,Maselli-2013,Yagi-2015,Yagi-2017,Alexander-2019,Martinon-2014,Brew-2016,Chatterjee-2025,Jyothilakshmi-2025,Anderson-1998,Lau-2010,Chen-2014,Chirenti,Lioutas-2021,Zhao-2022,Rodriguez-2025,Lattimer-2001,Reed-2020}, hot neutron stars~\cite{,Laskos-2022,Marques-2017,Raduta-2020,Khadkikar-2021} and compact stars involving exotic forms of matter~\cite{Largani-2022,Ranea-2023,Ranea-2022,Ranea-2022b}}. The most widely known universal relations are the so-called I-Love-Q relations which connect, in a remarkable EOS independent way, rescaled quantities involving the moment of inertia, the tidal deformability and the quadrupole moment~\cite{Yagi-2013}. In addition, the well-studied Love-C~\cite{Maselli-2013} relation is known to connect the tidal deformation to the compactness of a star and can be utilized to translate information from binary neutron star mergers to mass and radius constraints~\cite{Abbott-2018}. Interestingly, and maybe more importantly, apart from the indirect measurement of different stellar properties, universal relations set the ground for probing the theory of gravity~\cite{Yagi-2013}.

Since their discovery~\cite{Abbott-2016}, gravitational radiation or gravitational waves (GWs) are considered to be a crucial probe of astrophysical and cosmological phenomena. Being the propagating perturbation of the metric tensor~\cite{Misner-1973}, GWs are sourced by several phenomena of dynamical perturbations occurring in astrophysical and cosmological structure. The spectrum of their emitted frequencies can be attributed to different physical sources, ranging from compact object dynamics to perturbations in stellar matter or in the early universe. Current detectors, (LIGO,Virgo, KARGA, GEO600) are sensitive in the band of 10Hz up to a few kHz~\cite{Schnabel-2025}. At the upper range of this band, measurements are expected to find GWs coupled with the $f$ and $p$ modes of non-radial oscillations of compact stars, providing direct information on the properties of ultra-dense matter. Interestingly, such modes may appear when violent cosmic phenomena, such as neutron star mergers~\cite{Pietri-2020} or supernovae~\cite{Abdikamalov-2021,Radice-2019}, occur.

Interestingly, several research studies have been focused on the proposition of quasi-universal relations involving the frequency of non-radial pulsations~\cite{Anderson-1998,Lau-2010,Chen-2014,Chirenti,Lioutas-2021,Zhao-2022,Rodriguez-2025}. Since the frequencies of these modes may be observationally attainable, they could provide important insight regarding other structural neutron star properties to which they are connected. Anderson and Kokkotas~\cite{Anderson-1998} were the first to show that the $f$ mode frequency $f_0=\omega_0/2\pi$ follows a quasi-universal relation with the mean density of a star $\sim\sqrt{M/R^3}$ ($M$ denotes the stellar mass and $R$ the radius). Later, it was found that the rescaled quantity $M\omega_0$ follows a series of more precise relations with the rescaled moment of inertia $\bar{I}$, the dimensionless tidal deformability $\Lambda$ and the compactness $C$~\cite{Lau-2010,Chen-2014,Chirenti,Lioutas-2021}. In ref.~\cite{Zhao-2022}, Zhao and Lattimer refined previously established universal relations involving the frequency of $f$, $p$ and $g$ modes by using modern EOSs that fulfill state-of-the-art multimessenger constraints.

Another important quantity that may be measurable in the event of a supernova explosion is the binding energy $E_b$ of a compact object~\cite{Lattimer-2007}. Being the difference between the baryonic and the gravitational mass the binding energy of a neutron star can be thought as an analogue of the nuclear binding energy. Interestingly, as suggested by Lattimer~\cite{Lattimer-2007}, the neutrinos emitted during supernovae carry information that may allow us to determine the neutron star binding energy. Once again, there have been investigations regarding the involvement of $E_b$ in universal {relations}~\cite{Lattimer-2001,Reed-2020,Laskos-2022,Brew-2016}. Lattimer and Prakash indicated an empirical connection between the binding energy and the stellar compactness~\cite{Lattimer-2001}, while Reed and {Horowitz}~\cite{Reed-2020} discussed a quasi-universal connection with the dimensionless tidal deformability. {Notably, the work of ref.~\cite{Raduta-2020} has examined the validity of the relation between the binding energy and the compactness at finite temperature}. In ref.~\cite{Laskos-2022}, the authors proposed a very precise universal relation between the binding energy, the {gravitational} redshift and the dimensionless tidal deformability that has an error of less than $0.6\%$ and also holds for finite temperature (isentropic) EOSs.

Given that both the binding energy and the frequency of a non-radial oscillation are involved in quasi-universal relations and that they may also be measured in the occasion of a core-collapse supernova explosion~\cite{Abdikamalov-2021,Lattimer-2007}, in the present study we aim to investigate if there is a possible empirical relation that connects them in a EOS independent way.

In order to extend our study, we do not only consider hadronic EOSs, but we also employ hybrid models featuring a first-order phase transition. {In particular, we utilize the well-known constant speed of sound method to construct a set of hybrid EOSs~\cite{Zdunik-2013,Alford-2013}. Notably, recent works on the subject have indicated that the occurrence of a phase transition can lead to deviations from universal relations~\cite{Ranea-2023,Ranea-2022,Ranea-2022b}. }

The importance of this study is threefold: Firstly, the existence of an empirical relation between the binding energy and the non-radial mode frequencies could allow us to probe the theory of gravity in a single event by utilizing multimessenger detection. Secondly, possible theoretical deviations from such an empirical relation based on the nature of the EOS could potentially highlight the existence of exotic forms of matter in compact stars. Last, any future insight on the binding energy could narrow down the range of possible frequencies for different modes, potentially aiding their detection.

The present paper is organized as follows: Section~{\ref{Xsec2-2}} contains the theoretical framework for the evaluation of the compact star binding energy, while Section~{\ref{Xsec3-3}} provides a detailed presentation of the formalism related to the non-radial oscillations. Section~{\ref{Xsec4-4}} collects the set of different EOSs employed in this study. In Section~{\ref{Xsec5-5}} we discuss our findings and their implications. Finally, Section~{\ref{Xsec6-6}} contains a summary of our results.

\section{Binding energy of compact stars}\label{Xsec2-2}\label{sec2}
The binding energy stands for the energy gain due to the assembling of $N$ baryons to form a stable star. As in the case of an atomic nucleus, the binding energy can be obtained via \cite{Lattimer-2001}  ($G=c=1$)
\begin{equation} \label{eq5}
    E_b=M_b-M=N m_b -M,
\end{equation}
where $m_b$ is the mass of single baryon, $M_b$ corresponds to the baryonic mass of the star and $M$ corresponds to the gravitational mass (to which we will refer simply as mass from now on) which is derived by the solution of Einstein's equations (or more precisely the TOV equations). Specifically,
\begin{equation} \label{eq6}
    M = 4\pi \int_0^R r^2\mathcal{E} (r)dr,
\end{equation}
where $\mathcal{E}(r)$ corresponds to the energy density distribution and $R$ is the radius of the star. The total number of nucleons $N$ is found by integrating the baryon density profile as
\begin{equation} \label{eq7}
    N=\int n(r) dV,
\end{equation}
where
\begin{equation} \label{eq8}
    dV =4 \pi r^2 \sqrt{g_{rr}(r)} dr=4 \pi r^2 e^{\Lambda(r)} dr,
\end{equation}
is the infinitesimal proper volume and
\begin{equation} \label{eq9}
    e^{\Lambda(r)}=\sqrt{g_{rr}(r)}=\left(1-\frac{2m(r)}{r}\right)^{-1/2}.
\end{equation}
In Eq.~{{(\ref{eq9})}}, $m(r)$ stands for the gravitational mass distribution of the star.

In the present study, following the authors of refs. \cite{Lattimer-2001,Reed-2020}, $m_b$ is taken to be the mass of $^{56}$Fe/$56=930.412$ MeV/$c^2$. Such choice is appropriate in the case where $E_b$ stands for the energy released during the core-collapse supernova of a white-dward-like iron core~\cite{Lattimer-2001}. Interestingly, the binding energy can be decuded by the detection of neutrinos from a supernova event. As indicated by Lattimer and Prakash~\cite{Lattimer-2001,Lattimer-2007}, this quantity might be the most precisely determined aspect of the respective neutrino signal.

\section{Non-radial stellar oscillations}\label{Xsec3-3}\label{sec3}

{The framework for the study of non-radial stellar oscillations {in General Relativity} was given in the pioneering study of Thorne and Campolattaro~\cite{Thorne-1967}. Several years later, Lindblom and Detweiler~\cite{Lindblom-1983} managed to reduce the {relevant} equations to a numerically solvable system and studied the quadrupole ($l = 2$) oscillations of neutron stars. This system of coupled differential equations, given the appropriate boundary conditions, forms a Sturm-Liouville eigenvalue problem for the oscillation frequency $\omega$.}

A solution of the system, describing non-radial pulsations, may be obtained by employing a weak coupling approximation between gravity and matter fields. This approach is known as Cowling approximation, named after T.G. Cowling who made a corresponding study on the non-radial pulsations of Newtonian stars in 1941  \cite{Cowling1941}. The relativistic Cowling approximation was developed originally by McDermott {{\rm et al.}}~\cite{McDermott-1983} and modified later by Finn \cite{Finn1988}. A comparison concerning these two branches of the relativistic Cowling approximation was given in ref.~\cite{Lindblom-1990}.

For the calculation of oscillation modes, in the relativistic Cowling approximation, we follow the formalism described by Sotani {{\rm et al.}}~\cite{Sotani-2011}. All metric perturbations are set to zero, and  the local conservation of the energy momentum tensor perturbations is being solved. In this framework, given the spherically symmetric background metric (describing the unperturbed stellar configuration)
\begin{equation}\label{eq6}
     ds^2=-e^{2\Phi(r)}dt^2+e^{2\Lambda(r)}dr^2+r^2(d\theta^2+\sin^2{\theta d\phi^2)},
 \end{equation}
the differential equations that need to be solved for the extraction of different mode frequencies are given as~\cite{Sotani-2011}
\begin{equation}\label{eq7}
 \begin{aligned}
        W'(r) =\frac{d\mathcal{E}}{dP} & \Big(\omega^2r^2e^{\Lambda(r) -2\Phi(r)}V(r)  +\Phi'(r)W(r)\Big)\\
         &-l(l+1)e^{\Lambda(r)}V(r)
\end{aligned}
\end{equation}

\begin{equation}\label{eq8}
        V'(r)=2\Phi'(r)V(r)-e^{\Lambda(r)}\frac{W(r)}{r^2},
\end{equation}
where $W(r)$ and $V(r)$ denote the spatial part of the matter field perturbation parameters $W(r,t)$ and $V(r,t)$, for which a harmonic time dependence has been assumed.

In order to solve the Eqs.~{{(\ref{eq6})}} and (\ref{eq7}) Strum-Luville problem, one needs an appropriate set of boundary conditions. In particular, near the center of the star $W(r)$ and $V(r)$ have the form $W(r)=Cr^{l+1}$ and $V(r) = -Cr^l/l$, where $C$ is an arbitrary constant~\cite{Sotani-2011}.~At the surface of the star ($r=R$) the following condition needs to be met~\cite{Sotani-2011}
\begin{equation}\label{eq10}
    [\omega^2r^2e^{\Lambda-2\Phi}V+\Phi'W]_{r=R}=0.
\end{equation}
Notably, in the presence of a density discontinuity (due to a phase transition) an additional junction condition is required for one to solve the aforementioned differential equations. More precisely, at the interface of the discontinuity~\cite{Sotani-2011}
\begin{equation}\label{eq11}
    W_+ = W_-
\end{equation}
\begin{equation}\label{eq11}
    V_+=\frac{e^{2\Phi-\Lambda}}{\omega^2R_t^2}\left(\frac{\mathcal{E}_-+P}{\mathcal{E}_++P}[\omega^2R_t^2e^{\Lambda-2\Phi} V_-+\Phi'W_-]-\Phi'W_+\right),
\end{equation}
where $R_t$ is the radius at which the discontinuity appears and the signs $+,-$ indicate the function values at each side of the boundary between the two phases.

\begin{figure*}
\centerline{\includegraphics[width=\linewidth,scale=0.5]{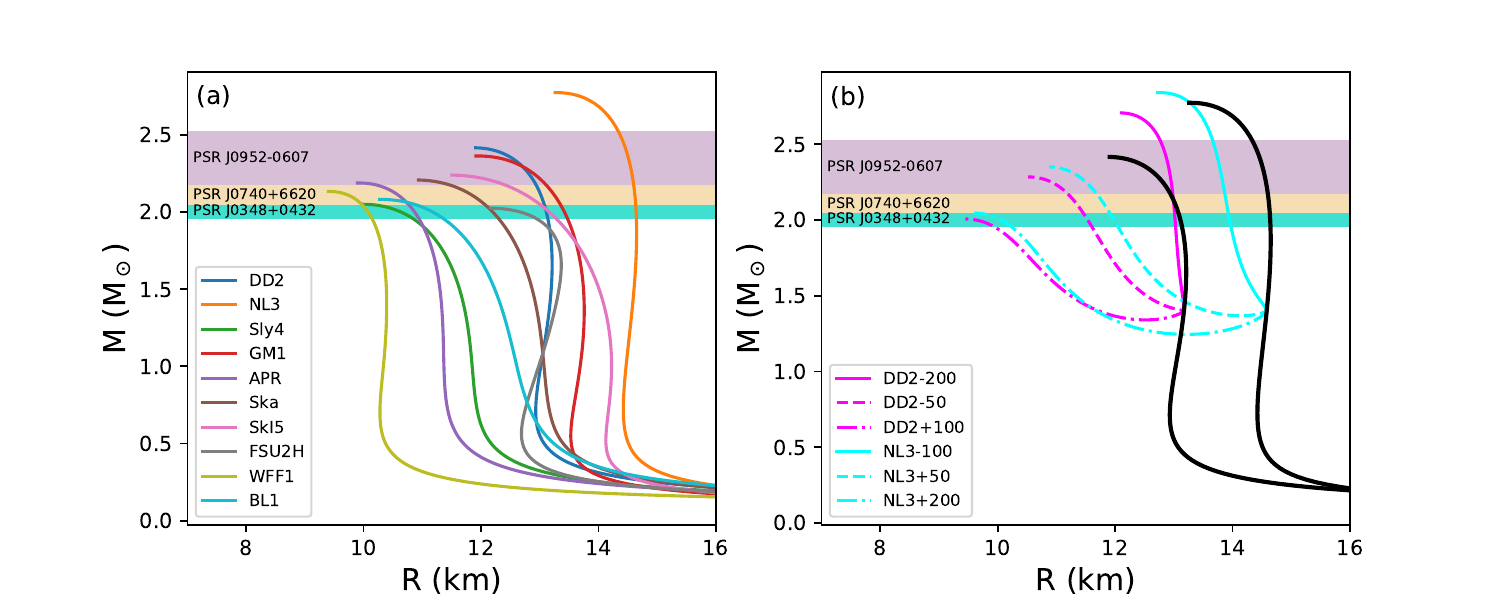}}
\caption{(a) Mass-radius diagrams for the hadronic EOSs employed in this study. (b) Mass-radius diagrams for the hybrid EOSs employed in this study. The number appearing in the legend indicates the difference between the energy density jump of the EOS and the critical energy density jump defined by Eq.~(\ref{eq13}). The
  shaded regions (in both panels) correspond to possible constraints on the maximum mass from the observation of PSR J0348\,+\,0432~\cite{Antoniadis-2013}, PSR J0740\,+\,6620~\cite{Cromatie-2020}, and PSR J0952-0607~\cite{Romani-2022}.}
\label{f1}
\end{figure*}

\begin{figure*}
\centerline{\includegraphics[width=\linewidth,scale=0.5]{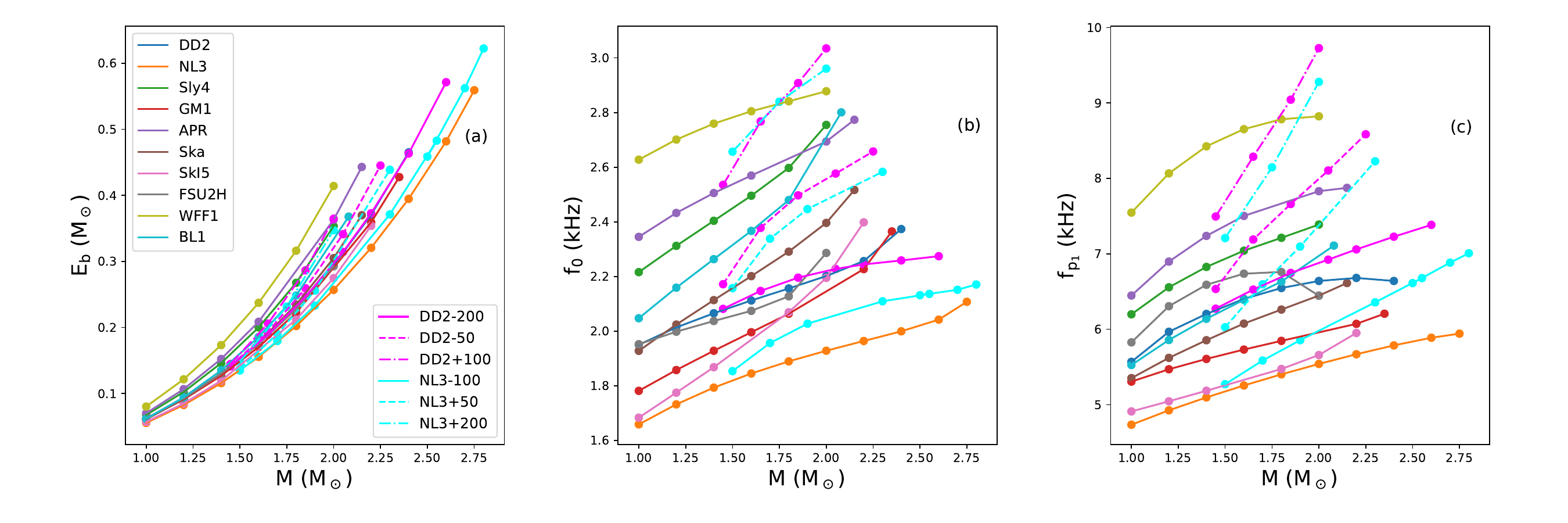}}
\caption{(a) The binding energy as a function of the gravitational mass for all of the employed EOSs. (b) The frequency of the $f$ mode oscillation as a function of a gravitational mass for all of the employed EOS. (c) Same as panel (b) but for $p_1$ modes.}
\label{f2}\vspace*{6pt}
\end{figure*}

\begin{figure*}
\centerline{\includegraphics[width=10 cm,scale=0.5]{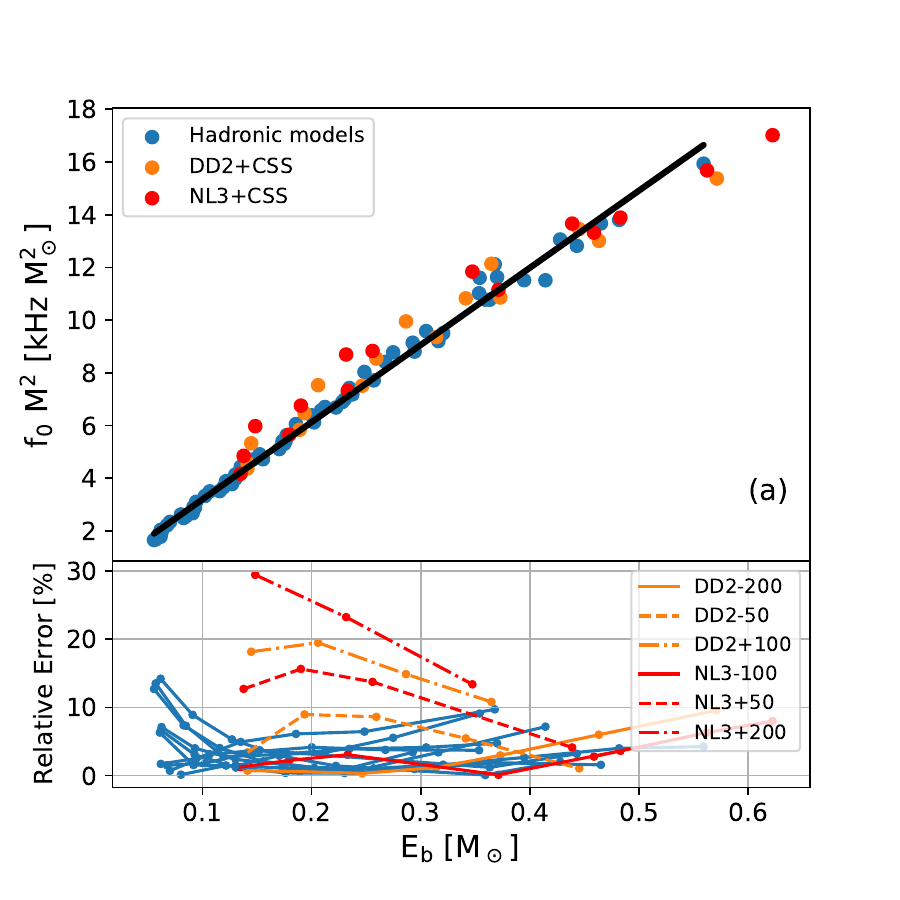}\includegraphics[width=10 cm,scale=0.5]{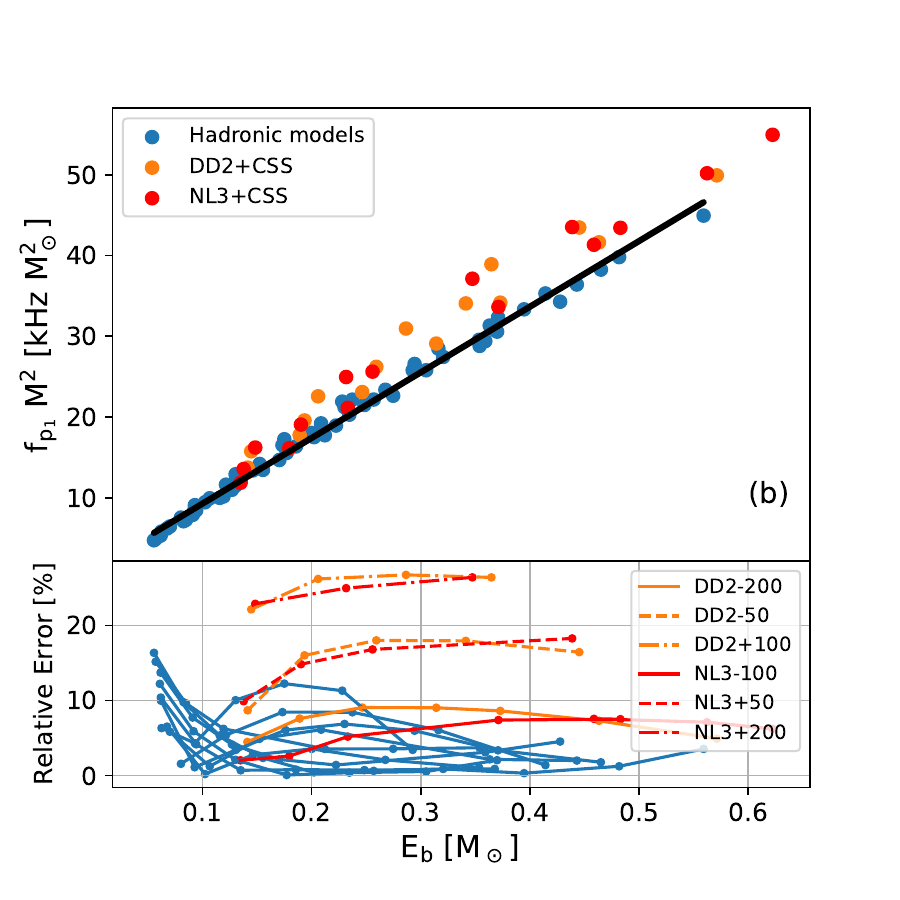}}
\caption{(a) The product of the $f$ mode oscillation frequency and the gravitational mass squared as a function of the binding energy appears in the top panel. The bottom panel includes the relative error ($100|y_{real}-y_{fit}|/y_{fit}$) from the fitted formula. (b) Same as panel (a) but the frequency is related to the $p_1$ mode oscillations. Note that for both panels the blue dots indicate results for hadronic EOSs, while the orange (red) dots dnote results for hybrid EOS constructed with the DD2 (NL3) model.}
\label{f3}
\end{figure*}

\begin{figure}
\centerline{\includegraphics[width=\linewidth,scale=0.5]{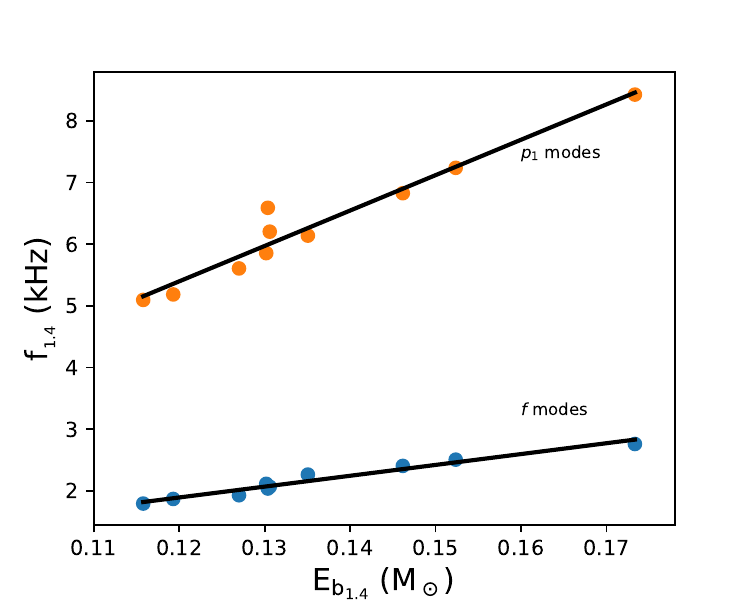}}
\caption{The frequency of $f$ and $p_1$ modes as a function of the binding energy for $1.4M_\odot$ hadronic stars.}
\label{f4}
\end{figure}

\section{Equations of state}\label{Xsec4-4}\label{sec4}
In the present study, we employed two distinct sets of EOSs. Specifically, we used a wide range of hadronic models to identify whether a correlation between the binding energy and the non-radial frequencies actually exists. {This variety of EOSs included relativistic mean-field (GRDF-DD2~\cite{Typel-2018}, NL3~\cite{Lalazissis-1997,Xia-2022a,Xia-2022b,Niu-2025}, GM1~\cite{Glendenning-1991,Douchin-2001}, FSU2H~\cite{Tolos-2017}), Skyrme (Sly4~\cite{Gulminelli-2015,Danielewicz-2009,Chabanat-1998}, Ska~\cite{Gulminelli-2015,Danielewicz-2009,Kohler-1976}, SkI5~\cite{Gulminelli-2015,Danielewicz-2009,Reinhard-1995}), and microscopic models (APR~\cite{Akmal-1998,Douchin-2001,Baym-1971}, BL~\cite{Bombaci-2018}, WFF1~\cite{Wiringa-1988}).~Note that the CompOSE repository~\cite{Typel-2015,Oertel-2017,Typel-2022,compose} has been used in the attempt of retrieving EOSs}.

{Furthermore}, we used a set of hybrid EOSs, incorporating a first-order phase transition, by combining two of the employed hadronic models with the well-known constant speed of sound model (CSS)~\cite{Zdunik-2013,Alford-2013}. The main advantage of working within the CSS approach is its flexibility. As this model is parametrized via the transition pressure, the energy density jump and the speed of sound, one can systematically assess whether a quasi-universal correlation would hold for hybrid models.

{Notably, following the approach of several related works~\cite{Christian-2019,Christian-2021,Christian-2022,Han-2019a,Sharifi-2021,Paschalidis-2018,Alford-2017,Han-2020}, the speed of sound parameter was set equal to the speed of light $c$. It is worth mentioning that, according to perturbative QCD demands, the speed of sound in quark matter should approach $c/\sqrt{3}$ as the density increases~\cite{Kurkela-2010,Hebeler-2013,Kurkela-2014}. However, the applicability of this constraint concerns a density range close to $\sim40n_0$ ($n_0$: saturation density), which is\break considered to be well-above the central density of compact stars~\cite{Tan-2022b}. In that sense the speed of sound should be reduced as the density further increases to satisfy such a demand (either continuously~\cite{Tews-2018} or potentially via a second QCD phase transition~\cite{Alford-2017}). For a comprehensive analysis on the speed of sound structure of hybrid EOSs, the reader is referred to ref.~\cite{Tan-2022b}.}

It is worth noting that several studies, working on universal relations, are only considering EOSs that are compatible to modern astronomical constraints~\cite{Zhao-2022,Brew-2016}. This is a reasonable approach as having a set of EOSs for which the predicted stellar properties form a narrower set may produce more accurate EOS-independent relations. In the present work, the only demand for the selected EOSs is that they are compatible to the existence of stable massive stars that exceed $2M_\odot$~\cite{Antoniadis-2013}. Not imposing modern radius constraints (e.g., GW170817/NICER etc.~\citet{Abbott-2018,Miller-2019,Riley-2019}) provides a more wide set of EOSs challenging the possible postulation of an empirical relation.

\section{Results}\label{Xsec5-5}\label{sec5}
To begin our analysis, we used a set of ten hadronic EOSs and six hybrid ones. In Fig.~{\ref{f1}} one can observe the mass-radius diagram predicted by the employed EOSs. As is evident, all of the employed models are in accordance to the $2M_\odot$ maximum mass constraint. In the case of hybrid EOSs, the parametrization has been achieved via the CSS model. For all of the constructed models, the phase transition occurs when the hadronic branch (in the $M-R$ diagram) reaches $1.4 M_\odot$ and the speed of sound $c_s$ is set equal to the speed of light. Hence, the only parameter that is varied is the energy density jump $\Delta\mathcal{E}$. Notably, the numbers appearing in the legend of Fig.~{\ref{f1}}(b) denote the difference between the selected energy density jump and the so-called critical energy density $\Delta\mathcal{E}_\mathrm{cr}$~\cite{Alford-2013}. This critical value for the discontinuity sets an approximate benchmark for the appearance of a descending branch in the mass-radius graph and it is given by~\cite{Alford-2013}
\begin{equation}\label{eq13}
    \Delta\mathcal{E}_\mathrm{cr}=\frac{1}{2}\mathcal{E}_\mathrm{tr}+\frac{3}{2}P_\mathrm{tr}.
\end{equation}
In Eq.~{(\ref{eq13})}, $\mathcal{E}_\mathrm{tr}$ and $P_\mathrm{tr}$ denote the transition energy density and pressure, respectively.

In Fig.~{\ref{f2}}(a), we plotted the binding energy as a function of the gravitational mass for all of the employed EOSs. As it is clear, stiffer EOSs lead to lower binding energy values for a given mass. This could be related to the fact that compact stars of specific mass exhibit lower central baryon densities when constructed with stiffer models. In Fig.~{\ref{f2}}(b) and (c) one can find the oscillation frequency as a function of the mass for $f$ and $p_1$ modes, respectively. Interestingly, softer models tend to predict higher frequency value for a given mass. Note that the dots appearing in Fig.~{\ref{f2}} denote the configurations for which we investigated the existence of a possible correlation between $\omega$ and $E_b$. As one can observe, the employed models scan a wide range of possible oscillation frequencies and binding energies, allowing us to proceed with our analysis.

Fig.~{\ref{f3}}(a) depicts the dependence between the binding energy and the $f$ mode frequency multiplied by the gravitational mass squared for all of the constructed stellar configurations. The blue dots denote the predictions related to hadronic configurations, while the orange and red points stand for hybrid models (constructed with the DD2 and the NL3 model, respectively). As one can observe, the hadronic configurations appear to follow a rather precise and nearly universal trend. For the latter reason, we performed linear regression to the corresponding data of the following form
\begin{equation}\label{fit1}
    \frac{f_0M^2}{\mathrm{kHz} M_\odot^2} = 0.27+29.29\frac{E_b}{M_\odot}.
\end{equation}
Notably, the correlation coefficient $R_{fit}^2$ was found to be equal to 0.99. In addition, the aforementioned formula reproduced the exact numerical data with an error lower than $10\%$ for compact stars with masses larger than $1.2M_\odot$.~As is also evident the consideration of a phase transition in dense matter may lead to significant deviations from the proposed empirical relation in Eq.~{(\ref{fit1})}. In particular, we found that hybrid EOSs with low enegy density jump values exhibit deviations from Eq.~{(\ref{fit1})} that are similar to those related to purely hadronic EOSs. However, as the density discontinuity increases the deviations may reach values up to $30\%$.

Fig.~{\ref{f3}}(b) contains the same results as Fig.~{\ref{f3}}(a), but in this case the frequency value corresponds to $p_1$ mode oscillations. Notably, a relation of the form
\begin{equation}\label{fit2}
    \frac{f_{p_1}M^2}{\mathrm{kHz}M_\odot^2} = 1.13+81.30\frac{E_b}{M_\odot}.
\end{equation}
was also fitted and was found to reproduce the theoretical predictions with an accuracy similar to the fit applied for the $f$ modes (the error remains less than $\sim12\%$ for $M\geq1.2M_\odot$).~Another interesting remark is that, in the case of $f$ modes, the hybrid configurations that were found to exhibit the largest deviations from the fit were those close to the canonical neutron star mass ($\sim 1.4 M_\odot$). For larger values of masses the relative error has a decreasing trend. In contrast, in the case related to the $p_1$ mode the relative error reaches a plateau and remains there as the mass further increases.

As the vast majority of the observed compact stars have a mass close to the canonical one, we found important to also calculate the relation that connects the oscillation frequency and the binding energy for $1.4 M_\odot$ hadronic stars. The respective results can be found in Fig.~{\ref{f4}}. For the case of $f$ mode oscillations we found a formula of the form
\begin{equation}\label{fit3}
    \frac{f_0^{1.4}}{\mathrm{kHz}} = -0.22 + 17.60\frac{E_b^{1.4}}{M_\odot}.
\end{equation}
For the $p_1$ modes the corresponding correlation can be modeled via
\begin{equation}\label{fit4}
    f_{p_1}^{1.4} = -1.49+ 57.43E_b^{1.4}.
\end{equation}
Note that the superscript appearing in Eqs.~{(\ref{fit3})} and (\ref{fit4}) does not stand for a power but for an index (indicating $1.4M_\odot$). Once again, the error of the fitting formulas remains less than $10\%$ for both $f$ and $p_1$ modes.

\section{Summary and future directions}\label{Xsec6-6}\label{sec6}
In the present study, we investigated the possible existence of a correlation between the binding energy of compact stars and the frequencies of their non-radial pulsation modes. The analysis was performed within the relativistic Cowling approximation and it included the calculation of $f$ and $p_1$ modes. Notably, as core-collapse supernovae may be informative for both the binding energy and the oscillation frequencies, a single event could, in principle, provide important insight for the physics of compact stars.

Interestingly, we found that, for hadronic models, the $f$ mode frequency is correlated to the binding energy with a respective correlation $R_{fit}^2$ coefficient of 0.99. The case was similar for the $p_1$ modes as well. The derived linear formulas describing the dependence were found to reproduce the exact theoretical predictions within an error margin of $\sim10\%$ for $f$ modes and $\sim12\%$ for $p_1$ modes (for stellar configurations with masses larger than $1.2M_\odot$).

Another important remark is that the predictions of hybrid EOSs were found to deviate from the proposed fitted formulas, with the differences increasing for higher energy density jump values. Notably, in the case of $f$ modes, deviations close to $30\%$ were reported. As a consequence, the empirical formulas derived in the present work may potentially allow us to probe the nature of compact star cores. More\break precisely, significant deviations (of future precise astronomical measurements) from the predictions of such empirical relations could potentially be attributed to the existence of exotic degrees of freedom in the stellar interior.

Finally, it is crucial to clarify that this study is only a first step towards postulating an empirical relation between the binding energy and the frequencies of non-radial oscillations. While the Cowling approximation allows us to probe the potential existence of such a nearly universal connection, it is also important to perform the present computations {without employing the aforementioned simplification
in order to avoid
any errors related to the neglect of metric perturbations. Furthermore, for the hybrid models we restricted to a low transition density value close to $2n_0$. It would be potentially fruitful to also vary this parameter and assess its impact on the derived deviations from the empirical relation (for a recent study investigating the role of the transition density in light of recent data the reader is referred to ref.~\cite{Li-2023n})}. Another interesting research avenue would be to consider the possible effects of finite temperature on the selected models. In any case, the present results may provide a framework for further probing of compact star physics in the era of multimessenger astronomy.

\section*{Acknowledgements}
P.L.-P. acknowledges that the research work was supported by the Hellenic Foundation for Research and Innovation (HFRI) under the 5th call for HFRI PhD fellowships (Fellowship Number: 19175).



\end{document}